# Performance Improvement by Changing Modulation Methods for Software Defined Radios

Bhalchandra B. Godbole

Karmaveer Bhaurao Patil College of Engineering and
Polytechnic, Satara - 415001

bbgodbole@rediffmail.com

Dilip S. Aldar

Karmaveer Bhaurao Patil College of Engineering and
Polytechnic, Satara - 415001

dilip_aldar@rediffmail.com

*Abstract-* **This paper describes an automatic switching of modulation method to reconfigure transceivers of Software Defined Radio (SDR) based wireless communication system. The programmable architecture of Software Radio promotes a flexible implementation of modulation methods. This flexibility also translates into adaptively, which is used here to optimize the throughput of a wireless network, operating under varying channel conditions.**
**It is robust and efficient with processing time overhead that still allows the SDR to maintain its real-time operating objectives. This technique is studied for digital wireless communication systems.**
**Tests and simulations using an AWGN channel show that the SNR threshold is 5dB for the case study.**

## I. INTRODUCTION

Reconfiguration, dynamic/static, partial/complete is an essential part of software radio technology. Thanks to SDR, so that the systems can be designed for change and evolution. In other words "*change*" becomes part of mainstream system operation. Recent work in European, Asian, and American R&D projects and the SDR Forum, lately WWRF, clearly shows that the concept of reconfigurability especially in the context of mobile cellular networks is not very complicated business. Reconfiguration still raises question on required *system-level support* both at the reconfigured devices and the network side [1], [2].

Over the past decade, previous work has correctly demonstrated the technical feasibility of the SDR approach in the design of radio equipment. This potential can be concretely exploited through equipment reconfiguration. In the evolution path towards 4G and beyond, this potential can be useful in many technically challenging as well as commercially attractive scenarios. Optimization in the QoS will necessitate considering reconfiguration as part of the mainstream operation [3].

This paper describes one application that exploits the flexibility of a software radio. The ability to select automatically, the correct modulation scheme used in an unknown received signal is a major advantage in a wireless network. As a channel capacity varies, modulation scheme switching enables the baud rate to be increased or decreased thus maximizing channel capacity usage. However the finite processor computing power limits the complexity of software radio if real time constraints are to be met [4], [5]. This paper proposes an automatic modulation scheme-switching algorithm based on SDR for *General-Purpose Processor* (GPP), *Digital Signal Processing* (DSP) and /or VLSI chips like FPGA. In this paper, we have simulated the algorithm on GPP.

The GPP approach makes use of P-IV system for rapid application development, large amount of program memory, and relatively inexpensive compared to the inflexible dedicated hardware. Conventional DSP processors rely on assembly language optimization for maximization of application efficiency, but with the P-IV, the degree of optimization possible using high-level languages such as C++ and MATLAB is much greater [6]. Software Defined Radio system closely tracks the advances in new high-speed processor technology allowing the addition of more complex signal processing techniques to the SDR system, while maintaining the real-time objective.

While switching the modulation method, it should be ensured that both transmitter and receiver are in tandem. To do so, any of two existing techniques can be used:

1. Automatic Modulation Recognition Technique,
2. Cross Layer Adaptation Technique.

Work in automatic modulation recognition is interesting and extensive work has been carried out for a number of years producing processor intensive techniques mainly restricted to non real-time operation. Recently published modulation recognition algorithms include a decision theoretic approach and pattern recognition approach used to discriminate between digitally modulated signals [7]. Modulation recognition scheme by the signal envelope extraction method and a digital modulation scheme classifier based on a pattern recognition technique generalises the moment matrix technique to gray scale images used in binary image word-spotting problems. A successful algorithm for an SDR implementation must be robust and efficient but the processing overhead must not stop the software radio from maintaining its real-time operating objectives. As the techniques described above are in general very processor intensive, they are not presently suitable for the SDR systems [8]. Here, we are using the physical and MAC layer signaling and control channel for the transmitter and receiver handshaking to signal the switched modulation. But it is also possible to use modulation recognition instead of cross layer signaling and use synchronous communication without adding delays.

With an attempt to introduce briefly Software Defined Radios in Section I, we explain SDR terminal hardware and software architecture in Section II. Section III discusses Cross Layer Adaptation technique to handshake the modulation method switching. Section IV is on the Case Study in details with the algorithm for modulation switching and the simulation results are presented in Section V. The work is concluded in Section VI.

## II. SDR TERMINAL: ARCHITECTURE AND CONFIGURATION

### A. Hardware Architecture:

The future mobile terminals are designed to support high bit traffic and they will be certainly equipped with high computational capabilities [4], [9]. Fourth generation mobile terminals, as shown in figure-1, will use more than one processor to increase the global computational capacity of the device. The Operating System (OS) performs all the control functions. In a SDR mobile terminal, the OS can also be used to generate the firmware to be inserted into the reconfigurable strata of the hardware.

A feasible architecture of baseband digital signal processing section is constituted by the following blocks:

- A general purpose microprocessor;

- Multiple DSP/FPGA blocks, depending on functional requirements of device;

- FIFO memories for both the RX and TX branches;

- ROM/FLASH memories, for storing the resident parts of the OS;

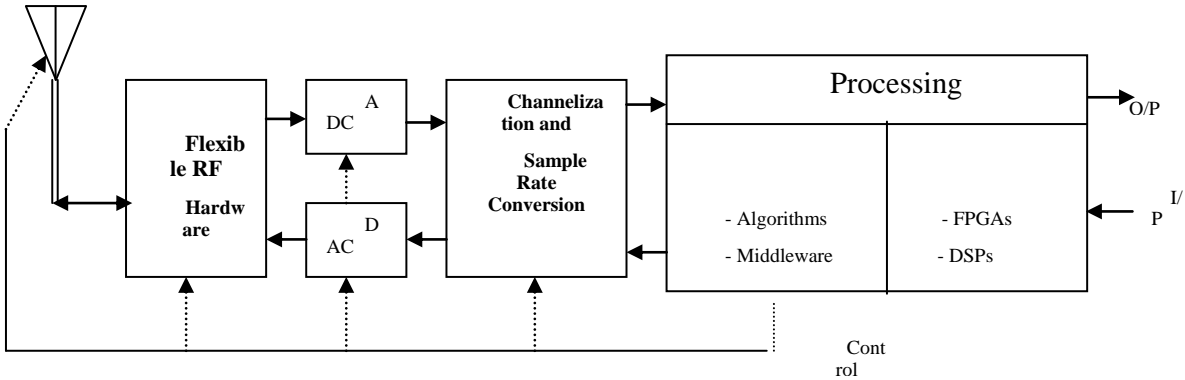

Figure 1. Model of a Software Radio

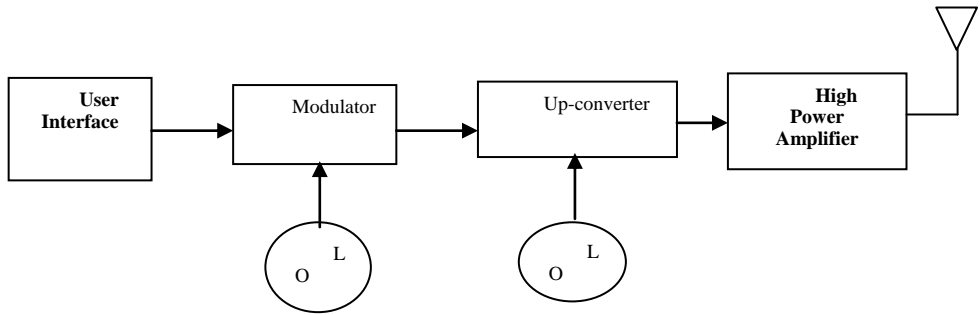

(a)Transmitter

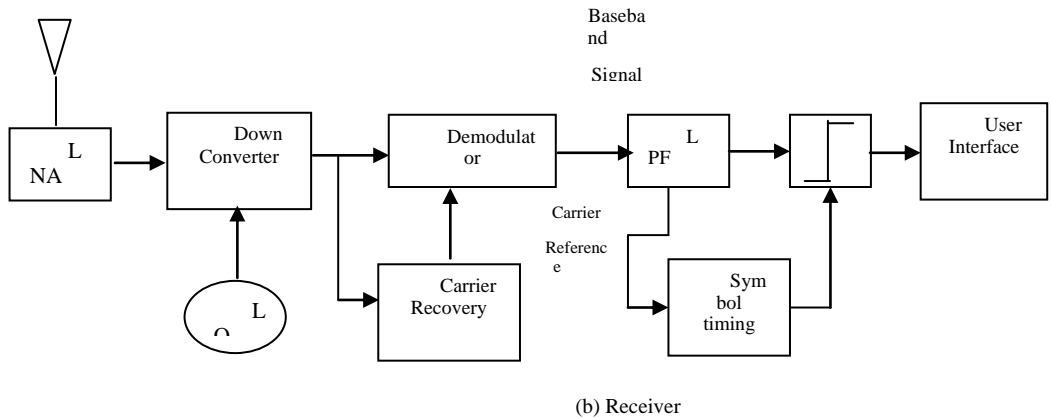

Figure-2. General Architecture of a transceiver in a software radio based wireless communication system.

The reconfigurable hardware, the FIFO Memories and the general purpose CPU are interconnected by high-speed bus. Alternately, the transceiver in a software radio based wireless communication system for GPP hardware will be as depicted in figure-2.

### B. Software Architecture

The device operating in SDR technology acts as a node in the hosting networks architecture both for control functions (control plane) and for the communication functions (traffic plane). The relevant entities of SDR network architecture are the configuration manager and the SDR mobile terminal. The physical layer of the mobile terminal is split into two sub-layers: the reconfigurable hardware and the macrocode (firmware), which implements the target communication protocol.

The key process for remote terminal configuration is the radio software download. The download process is constituted by phases like, pre-download, download and post-download [10]. The radio download data is intended platform dependent. In [11] the remote physical layer definition is operated through a high level descriptive language, called RADL (Radio Access Definition Language). As an example, functions like digital filtering, coding and decoding, modulation, pulse shaping, carrier recovery, timing acquisition are defined by a set of parameters provided by the SDR terminal using its libraries.

We simulate these functions using MATLAB and perform the sequence of operations as per requirement.

### III. CROSS LAYER APPROACH

Any innovation adopted at physical layer will not obtain the maximum performance if it does not interact with the upper MAC layer. This concept, opposed to the historical separation of functions between layers, has demonstrated significant results in terms of efficiency in the use of the communication resources. Cross-layer design, whichever is its target, requires a physical layer in communication with MAC not only giving information about the physical status of the channel, but also accepting real-time reconfiguration of transmission parameters. Also in this case, the SDR acts as the enabling technology, providing a full logical control over the signal processing functions located at physical layer.

The so-called cross-layer approach, depicted in figure-3, is the mutual exchange of information between physical & upper layers and is naturally enabled by Software Radio [12]. Several objects were proposed for cross-layer design. Some research has addressed improving QoS improvement and power management, proposing strategies based on modifying different layers of the communication system and also show that adapting transmission methodology to channel fading significantly improves link efficiency [13].

Therefore, assuming that all second and third generation systems can be simulated on the reconfigurable multi-processor platform, we propose modulation-switching algorithm to improve performance.

### IV. CASE STUDY

In order to achieve high data rate transmission under a target BER, to attain so-called Quality of Service (QoS), we propose modulation-switching algorithm to change the number of bits for each symbol. Computing receiver noise and SNR at a receiver is important for determining coverage and QoS in a wireless communication system. SNR determines the link quality and impacts the probability of error in a wireless communication system. Thus, the ability to estimate SNR is important for determining suitable transmitter powers or received signal levels in various propagation conditions. SNR at a receiver is a convenient metric that allows a designer to factor in the noise induced by the channel.

Modulation in channel has a significant effect on the quality of the information transfer measured in BER and on the complexity of the receiver. Receiver complexity dominates the complexity of SDR.

A receiver is typically four times more complex than a transmitter in terms of MIPS required to implement the baseband and the IF processing in the software. The digital modulation_demodulation algorithm topics include AGC,

Channel waveform coherence, coding/decoding and spreading /dispreading of the spectrum used [14]. In the present case study, we will be focusing on the modulation switching, therefore, AGC, coding and spreading are not discussed.

As shown in Figure-4, the probability of bit error is a function of channel modulation. There are broadly three main digital modulation types: ASK, FSK and PSK. To approach channel capacity upper bound, adaptive modulation with continuous modulation order have to be used. However, in reality, the continuous modulation order is very complex, therefore, discrete modulation orders, M = 2, 4, 16, 64 are used here, where M = 2 is the BPSK modulation, M = 4 is the

QPSK modulation, and 16, 64 are M-QAM modulations. When the full CSI is known at the transmitter side, it is possible to estimate the channel quality by SNRs. A channel with better quality can be assigned a larger number of bits and a higher order modulation, whereas a channel with poorer quality has to be assigned fewer bits or no bit when the channel quality is too bad to transmit even BPSK modulated signals.

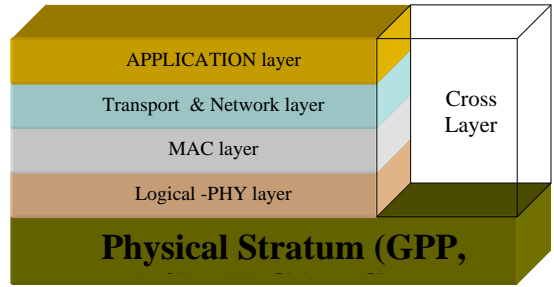

Figure-3 The cross-layer concept

1. Speech Coder =47.2 %

2. Channel Encoder, interleaver, cipher = 2.7 %

3. Speech Decoder= 16.8 %

4. Decipher & deinterleaver =0.73 %

5. Channel Decoder =12.0 %

6. Modulator = 13.6 %

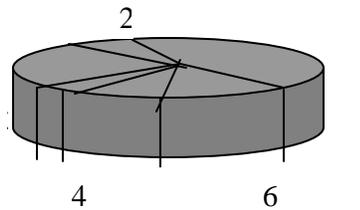

Figure-4 CPU requirement for EIGHT logical channels on Pentium-IV: 2 GHz [14].

For a certain modulation type, the relationship of modulation order M, SNR and baud, $P_b$ performance can be expressed as:

$$P_b = f(SNR, M, baud) \qquad (1)$$

Some of the F functions can be solved only numerically, but a number of them have a closed-form solution. For example, in flat Rayleigh fading channels, The BER for coherently detected M-QAM with Gray bit mapping is approximated. This paper presents an SR system physical layer, which is less complex, to meet constraints set by cellular mobile communication system depending on the specific application and type of data. A communication system must meet various QoS constraints e.g. BER, data rate, or energy dissipation at different times. For example, a cellular system may require a low BER for control data, a high data rate, or minimal energy dissipation for regular updates.

Typical communication systems are built to meet these QoS constraints ever under the worst calculation. However, a communication system operates in the worst condition infrequently. So, it wastes valuable resource as a result existing approaches to meet QoS requirements in cellular systems focus on MAC layer and above, but these optimization are often application dependent. Adaptive technique is 17 dB more power efficient than non-adaptive modulation in fading [11].

In the Physical Layer, the transmitter is the energy efficient pulse generator. The transmitter modulates an input bit stream into a wave of sinusoids with center frequency of 900 MHz and a bandwidth of 2.5 MHz.

$$S(t) = \sum_{i=-\infty}^{\infty} A\, p\,(t - iTs - \delta * di(t)) \qquad (2)$$

Where, $di(t)$ shifts the phase in time by some multiple $\delta$, which is larger than the Ts to assure an orthogonal signal set.

The channel model may include AWGN, multipath effects, and also sources of co-channel interference. But, the model under study is AWGN. Note that techniques such as coding and spreading would achieve better performance in a practical system, but they are omitted from the model to focus on the effects of the modulation scheme [12].

In relation to the modulation switching scheme, the function of the MAC protocol is inter-node communication of the current conditions, resources, and, QoS requirements. It communicates these directly to and from the Physical Layer as header information. The MAC is also responsible for tracking such Physical Layer conditions as channel conditions and link distances.

The Application Layer provides pre-defined QoS constraints depending on the data type. For example, control data may require the lowest possible BER because it is typically not resilient to errors. The error rate and data rate could be sacrificed to operate with minimal energy dissipation. Still other applications may desire a trade-off that results in a compromise between data rate, BER, and energy dissipation [13].

TABLE I. COMPARISON OF MODULATION METHODS WITH DIFFERENT DATA RATES

| Data Rate (Mbps) | Modulation | $N_d$ bps | Data transfer* Duration (µS) |
|---|---|---|---|
| 3 | BPSK | 24 | 2012 |
| 6 | QPSK | 48 | 1008 |
| 12 | 16-QAM | 96 | 504 |
| 24 | 64-QAM | 192 | 252 |
| *for the processor Pentium-IV: 2 GHz. | | | |

The overall strategy for choosing a modulation is motivated by the behavior of MQAM in different environmental conditions. This section characterizes some of this behavior. Configurations with high M and short Ts are more susceptible to interference, which eventually limits the performance regardless of the Eb/N0. In contrast, configurations with low *M* and along Ts are less susceptible to interference, and therefore the performance is more limited by the Eb/No ratio [14]. Therefore, for low Eb/No ratios, modulation schemes with high *M* are preferable. However, for higher Eb/No ratios, the interference dominates the noise, and modulation schemes with low values of *M* are preferable.

The *local cost function* ($\gamma$) defines the cost of a transaction.

$$\gamma = \frac{(\alpha \cdot BER) \cdot (\beta \cdot Eb/No)}{(\chi \cdot Data\ Rate)} \qquad (3)$$

In the local function, each performance parameter includes a weighting function that shows the relative contribution of BER, energy, and data rate to the overall cost. In (2), $\alpha$ □is the weighting function of the BER, $\beta$□ is the weighting function of the energy, and $\chi$ □is the weighting function of the data rate. The BER is a function $W$ of M, *Ts, Eb/No*, channel impulse response $h(x)$, and interference $i(x)$. The *Eb/No* is a function $Y$ of the transmitter-receiver distance, the channel impulse response, and the maximum radiated energy allowed by the ITU-R for a given data rate. The data rate depends only on the Ts and M.

$$BER = W\,(Eb/No, Ts, m, h(x), i(x)) \qquad (4)$$

$$Eb/No = Y\,(D, m, Ts, h(x)) \qquad (5)$$

$$Bit\ Rate = \log_2 M / T_s \qquad (6)$$

Thus, from (2) - (5), $\gamma(z)$ defines the local cost function, where $z$ is a set of environmental and QoS 6-tuples [Eb/N0, m, Ts, h(x), i(x), dist] within $Z$, the set of all possible 6-tuples.

$$\gamma(z) = \frac{\chi}{\alpha\beta} \cdot \frac{Ts \cdot W(Eb/No, Ts, m, h(x) \cdot i(x))\ Y(Distance, m, Ts, h(x))}{\log_2 m} \qquad (7)$$

When environmental conditions vary over several transactions, the local cost function is described in terms of the *local expected cost function*

$$\bar{\gamma} = \sum_{\forall z \in Z} \gamma(z) \cdot p(z) \qquad (8)$$

Where $p(z)$ is the probability distribution of $z \in$ □Z, and p(z) is determined from observation. The system chooses a modulation scheme by optimizing the local expected cost function. A configuration is said to be optimal if it costs less than any other possible configuration for the given environmental and preset parameters. Numerical gradient-based optimization techniques can be used to minimize or maximize the cost function. However, gradient calculations can be costly for implementation, and hence, calculation may consume more resources than it saves. To reduce implementation complexity, the nodes use linear approximations to choose the optimal modulation scheme for the current operating conditions [15], [16], [17].

## V. SIMULATION RESULTS

The first case is minimum BER, has a fixed data rate, and radiates the maximum allowed Eb. The distances of each transaction determine the Eb/No such that it varies from 5 dB to 12 dB in integer steps, and the distribution is shown in (9).

$$P(E_b/N_o) = \{0.0525, 0 dB \le E_b/No \le 25 dB$$
$$0 \quad else \tag{9}$$

The channel impulse response is a random instance of the model for each transaction, and considers M and Ts values that result in the minimum BER without regard without unacceptably high-energy dissipation or unacceptably low data rate.

Next, the system maximizes the data rate for a target BER in various channel conditions. The environmental and QoS parameters are the same as above, but now the system increases the data rate by choosing an appropriate modulation method. Any non-adaptive system must operate at the fixed data rate of 25 Mbps to meet the target BER even in the best channel conditions.

Third, the system minimizes the radiated energy over various QoS constraints. The first QoS constraint requires a BER of $5 \times 10^{-5}$ and a data rate of 3 Mbps, and this represents the case of distributing microcode over several hops. The second case requires a BER of $2 \times 10^{-4}$ and a data rate of 12 Mbps, and this represents the case of video data.

For different scenarios [17], [18] the system always performs best, and different fixed systems performed second best depending on the QoS goal. For example, the fixed system with $M=16$ achieved the second best BER when BER is the goal, and the fixed system with $M=4$ achieved the best energy efficiency when energy efficiency is goal. Thus, the modulation-switching algorithm should be especially suitable for systems that have changing QoS goals. For example, a sensor network may wish to minimize energy dissipation during normal operation, minimize BER for control data, and maximize data rate when it detects an event. We now compare the performance of the proposed system to the fixed modulation systems as all three of the QoS goals change.

The following results consider that the goals of minimum BER, minimum energy, and maximum data rate all occur with equal probability. Further, the network designer weights each performance parameter such that $\alpha = \beta = \chi = 1$. Note that designers are free to change the weighting function to result in optimal performance for specific applications. The results in Table-2 are normalized and obtained from the cost function of (8). The proposed system performs much better each of the fixed modulation systems. This is in contrast to the previous results that concentrate on just one QoS parameter. In the previous results, one system may have similar performance to the proposed system.

Simulations show that the algorithm improves performance significantly as compared to a conventional, fixed modulation system under variable environmental and QoS requirements. The proposed system improves BER up to 50%, data rate up to 100 %, without sacrificing performance of any other parameter. When the QoS goals change dynamically, the modulation switching system performs significantly better than for a static QoS goal. The proposed Cellular Mobile Software Radios system improves performance for ad hoc and sensor networks by supplementing the MAC protocols in the previous chapters to improve BER, energy efficiency, or data rate. Further, it does not increase the hardware cost or complexity of the Cellular Mobile Software radio transceiver architecture, because it requires changes only to the control logic [19]. Next, a cross-level optimization scheme adapts the Cellular Mobile Software radio to meet various application level QoS constraints as channel conditions change

Cross layer optimization further improves performance for any protocol. The protocols and the optimization scheme are custom tailored to meet the requirements of both Cellular Mobile Software radio. Thus, they significantly outperform more general approaches.

The cpu-time computed using profiler with MATLAB-7 shows that the time required for fixed modulation and that for the modulation switching is approximately same. Also, the data rate is improved. Therefore, on account of negligible complexity the proposed scheme of modulation switching to improve performance of mobile communication system is efficient.

## VI. CONCLUSION

Most of the technological efforts in the wireless communication area are devoted to increase the rational us of resources and devices with two main objectives:

- To increase of efficiency, measured in terms o radio coverage, number of served users, power consumption, spectrum usage, biological impact, short time-to-market and fast network (re)-planning,

- To provide a good degree of services for next generation systems, possibly with a contained investment.

The success of modulation switching algorithm integration into commercial standards such as 3G, WLAN, and beyond (4G, short range communications, etc.) will rely on a fine compromise between rate maximization (Layered type) and diversity (space-time coding) solutions, also including the ability to adapt to the time changing nature of the wireless channel using some form of feedback.

This paper proposes an adaptive system that adapts its resources to efficiently meet QoS requirements in dynamic channel conditions. The system is particularly suitable for, which have demanding QoS requirements that change for various types of data. The CA block works across the Application, MAC and Physical Layers, which is a departure from established network design techniques. Although this complicates the design process, it meets the special demands of Cellular Mobile Software Radios.

Finally upcoming trials and performance measurements in specific deployment conditions will be required in order to evaluate precisely the overall benefits of SDR systems in real-time world wireless scenarios (for the actual and future services).

TABLE II. BER OF FIXED MODULATION VERSUS MODULATION SWITCHING SYSTEM

| Data Rate (Mbps) | Fixed modulation BER | | | | | | Proposed System BER | Percentage Decrease in BER |
|---|---|---|---|---|---|---|---|---|
| | M=2 | M=4 | M=8 | M=16 | M=32 | M=64 | | |
| 12 | 0.0474 | 0.0310 | 0.0303 | 0.0289 | 0.0355 | 0.0365 | 0.0263 | 44.7% |
| 3 | 0.0418 | 0.0239 | 0.0233 | 0.0252 | 0.0322 | 0.0332 | 0.0212 | 49.3% |

TABLE III. DATA RATE OF FIXED MODULATION VERSUS MODULATION SWITCHING SYSTEM

| Fixed modulation System Data Rate | Modulation switching System Data Rate | Percentage Increase in Data Rate |
|---|---|---|
| 12Mbps | 18.5 Mbps | 68.2% |
| 3 Mbps | 6.1 Mbps | 100% |

.

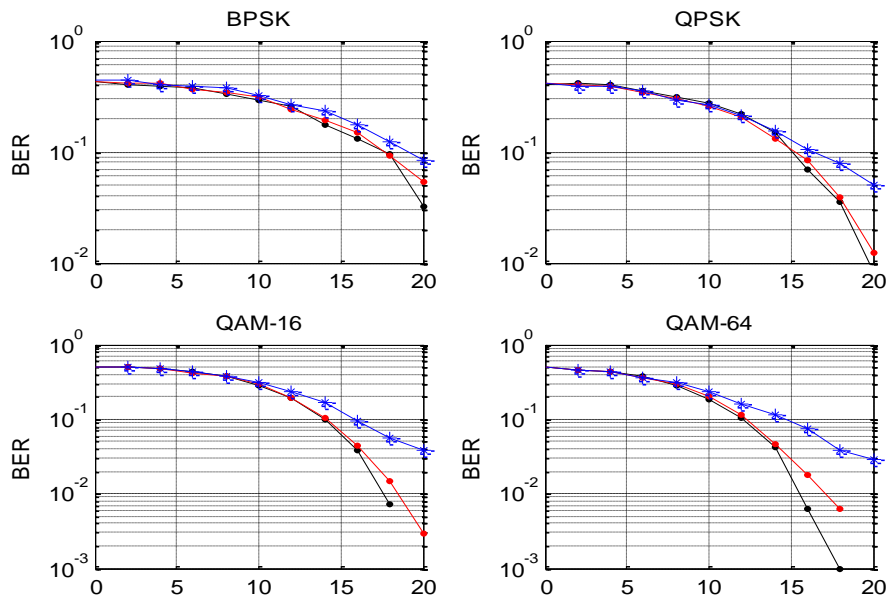

Figure-6. BER versus $E_b/No$ for modulation schemes used for simulation